\documentclass[a4paper,12pt]{article}
\usepackage{epsfig}
\usepackage{amssymb}
\usepackage{amsfonts}
\usepackage{amsmath}
\usepackage{euscript}
\usepackage{verbatim}
\usepackage{latexsym}
\usepackage{graphicx}

\newskip\humongous \humongous=0pt plus 1000pt minus 1000pt

\newif\ifdtup

\catcode`\@=11

\@addtoreset{equation}{section}

\def\@normalsize{\@setsize\normalsize{15pt}\xiipt\@xiipt
\abovedisplayskip 14pt plus3pt minus3pt%
\belowdisplayskip \abovedisplayskip
\abovedisplayshortskip \z@ plus3pt%
\belowdisplayshortskip 7pt plus3.5pt minus0pt}

\def\small{\@setsize\small{13.6pt}\xipt\@xipt
\abovedisplayskip 13pt plus3pt minus3pt%
\belowdisplayskip \abovedisplayskip
\abovedisplayshortskip \z@ plus3pt%
\belowdisplayshortskip 7pt plus3.5pt minus0pt
\def\@listi{\parsep 4.5pt plus 2pt minus 1pt
     \itemsep \parsep
     \topsep 9pt plus 3pt minus 3pt}}

\relax

\catcode`@=12

\catcode`\@=11

\def\section{\@startsection{section}{1}{\z@}{3.5ex plus 1ex minus
   .2ex}{2.3ex plus .2ex}{\large\bf}}

\def\SymBoxes#1#2#3#4{\newdimen\un@t \un@t#3%
\raisebox{#1}{\rule{#2\un@t}{#4}\hskip-#2\un@t% lower horizontal
\@tempdimb\un@t \advance\@tempdimb by-#4\@tempcntb#2\relax%
\@whilenum{\@tempcntb>0}\do{%                         % #2 vertical lines
\rule{#4}{\un@t}\hskip\@tempdimb \advance\@tempcntb by\m@ne}%
\hskip-#2\un@t \rule[\un@t]{#2\un@t}{#4}%
\rule[\un@t]{#4}{#4}\hskip-#4%             % upper horizontal line
\rule{#4}{\un@t}}\hskip-#4}                % rightest vertical line
\begin{document}
%\begin{letter}{~}

%%%%%%Define some new commands and  macros
\newcommand{\beq}{\begin{equation}}
\newcommand{\eeq}{\end{equation}}
\newcommand{\bea}{\begin{eqnarray}}
\newcommand{\eea}{\end{eqnarray}}
\newcommand{\beas}{\begin{eqnarray*}}
\newcommand{\eeas}{\end{eqnarray*}}
\newcommand{\defi}{\stackrel{\rm def}{=}}
\newcommand{\non}{\nonumber}
\newcommand{\bquo}{\begin{quote}}
\newcommand{\enqu}{\end{quote}}
%%%%%%%%%%%%%%%%
\renewcommand{\(}{\begin{equation}}
\renewcommand{\)}{\end{equation}}
%%%%%%%%%%%%%%%%%%%%%%%%%%%%%%%%%% definitions
\def \eqn#1#2{\begin{equation}#2\label{#1}\end{equation}}
\def\IZ{{\mathbb Z}}
\def\IR{{\mathbb R}}
\def\IC{{\mathbb C}}
\def\IQ{{\mathbb Q}}
\def\de{\partial}
\def\Tr{ \hbox{\rm Tr}}
\def\H{ \hbox{\rm H}}
\def\HE{ \hbox{$\rm H^{even}$}}
\def\HO{ \hbox{$\rm H^{odd}$}}
\def\K{ \hbox{\rm K}}
\def\Im{ \hbox{\rm Im}}
\def\Ker{ \hbox{\rm Ker}}
\def\const{\hbox {\rm const.}}
\def\o{\over}
\def\im{\hbox{\rm Im}}
\def\re{\hbox{\rm Re}}
\def\bra{\langle}\def\ket{\rangle}
\def\Arg{\hbox {\rm Arg}}
\def\Re{\hbox {\rm Re}}
\def\Im{\hbox {\rm Im}}
\def\exo{\hbox {\rm exp}}
\def\diag{\hbox{\rm diag}}
\def\longvert{{\rule[-2mm]{0.1mm}{7mm}}\,}
\def\a{\alpha}
\def\dag{{}^{\dagger}}
\def\tq{{\widetilde q}}
\def\p{{}^{\prime}}
\def\W{W}
\def\N{{\cal N}}
\def\hsp{,\hspace{.7cm}}

\def\br{\nonumber\\}
\def\IZ{{\mathbb Z}}
\def\IR{{\mathbb R}}
\def\IC{{\mathbb C}}
\def\IQ{{\mathbb Q}}
\def\IP{{\mathbb P}}
\def \eqn#1#2{\begin{equation}#2\label{#1}\end{equation}}

\newcommand{\C}{\ensuremath{\mathbb C}}
\newcommand{\Z}{\ensuremath{\mathbb Z}}
\newcommand{\R}{\ensuremath{\mathbb R}}
\newcommand{\rp}{\ensuremath{\mathbb {RP}}}
\newcommand{\cp}{\ensuremath{\mathbb {CP}}}
\newcommand{\vac}{\ensuremath{|0\rangle}}
\newcommand{\vact}{\ensuremath{|00\rangle}                    }
\newcommand{\oc}{\ensuremath{\overline{c}}}
\begin{titlepage}
\begin{flushright}
%CHEP XXXXX
%ULB-TH/09-10\\
%hep-th/yymmnnn\\
\end{flushright}
\bigskip
\def\thefootnote{\fnsymbol{footnote}}

\begin{center}
{\Large
{\bf Higher Spin Resolution of a Toy Big Bang
}
}
\end{center}

\bigskip
\begin{center}
{\large  Chethan KRISHNAN$^a$\footnote{\texttt{chethan@cts.iisc.ernet.in}} and   
Shubho ROY$^a$\footnote{{\texttt{sroy@het.brown.edu}}}
%Avinash RAJU$^a$\footnote{{\texttt{avinashraju777@gmail.com}}} \\
%\vspace{0.1in}
%and Somyadip THAKUR$^a$\footnote{{\texttt{smydp3thkr@gmail.com}}} 
}
\vspace{0.1in}

\end{center}

\renewcommand{\thefootnote}{\arabic{footnote}}

\begin{center}
%\vspace{0.2cm}
$^a$ {Center for High Energy Physics\\
Indian Institute of Science, Bangalore, India}\\

\end{center}

\noindent
\begin{center} {\bf Abstract} \end{center}
Diffeomorphisms preserve spacetime singularities, whereas higher spin symmetries need not. Since three dimensional de Sitter space has quotients that have big-bang/big-crunch singularities and since dS$_3$-gravity can be written as an $SL(2,\IC)$ Chern-Simons theory, we investigate $SL(3,\IC)$ Chern-Simons theory as a higher-spin context in which these singularities might get resolved. As in the case of higher spin black holes in $AdS_3$, the solutions are invariantly characterized by their holonomies. We show that the dS$_3$ quotient singularity can be de-singularized by an $SL(3,\IC)$ gauge transformation that preserves the holonomy: this is a higher spin resolution the cosmological singularity. 
%We also speculate about a possible connection between the finiteness of the Euclidean partition function (\'a la ``de Sitter Farey tail") and the resolvability of the singularity in the Lorentzian theory.
Our work deals exclusively with the bulk theory, and is independent of the subtleties involved in defining a CFT$_2$ dual to dS$_3$ in the sense of dS/CFT. 

\vspace{1.6 cm}
\vfill

\end{titlepage}

\setcounter{footnote}{0}

%%%%%%%%%%%%%%%%%%%%%%%%%%%%%%%%%%%%%%%%%%%%%%%%%%%%%%%%%%%%%%%%%%%%%%%%%%%%%%%%%%%%%%%%%%%%%%
%%%%%%%%%%%%%%%%%%%%%%%%%%%%%%%%%%%%%%%%%%%%%%%%%%%%%%%%%%%%%%%%%%%%%%%%%%%%%%%%%%%%%%%%%%%%%%
\section{Introduction}
\label{intro}

Spacetime singularities are one of the primary indications that general relativity should be modified at short distances. %\cite{Hawking-Penrose}. 
%by some UV complete (quantum) theory of gravity. 
String theory is an arena where this question can in principle be well-posed. Various examples of singularity resolution are known in string theory, eg. \cite{StromingerConifold}, \cite{Atish}\footnote{See \cite{KITP} for reviews.}. But it is fair to say that a systematic understanding of singularity resolution is still lacking. %\cite{BenCraps}.
%However, due to various technical and conceptual problems, progress in this direction has been painfully slow \cite{BenCraps}. 

Another question where the promise of string theory has not born substantial fruit is in the understanding of time-dependent backgrounds, aka cosmologies. (See \cite{Seiberg1, Seiberg2, Costa1, Costa2, Costa3} for various attempts in this direction.) This is related to the fact that typically, when it is under analytic sontrol, string theory is tied to supersymmetry. Unfortunately, supersymmetric backgrounds are necesarily time-independent and this frustrates most attempts to make progress on time-dependence in string theory. 

Together, the above two challenges imply that singularities in cosmological spaetimes are one of the hardest things to make sense of in the context of string theory\footnote{But see \cite{Ben1, Ben2, Evnin} for some progress in understanding light-like singularities in pp-waves and plane wave backgrounds using M(atrix)-theory \cite{Willy}.}.

On a different front, after the work of Vasiliev and others \cite{Vasiliev1, KP, Gop1}, a lot of recent atention has been directed towards an understanding of interacting higher spin theories. One motivation for this interest is the belief that the tensionless ($\alpha' \rightarrow \infty$) limit of string theory is a higher spin theory \cite{Sundborg}, and therefore higher spin theories might be a good starting point for a tractable understanding of (some) intrinsically stringy phenomena. Besides, we know that the dynamics of spin-1 fields is that of gauge vector fields living in a fixed spacetime background and that the dynamics of spin-2 fields gives rise to metric fluctuations and a dynamical spacetime. So perhaps it is not surprising that higher spin theories turn out to be relevant in our quest for a deeper understanding of the role of spacetime in string theory, ranging from background independence and singularity resolution to non-perturbative questions and the role of boundary conditions.%, perhaps it is not surprising that higher spin theories turn out to be relevant. 

% Our goal in this paper will be to show that a similar phenomenon exists also in a cosmological toy model in three dimensions, and that the cosmological singularity can be ``resolved" by a choice of gauge.

Vasiliev's intercating higher spin theories take their full glory in four and higher dimensions, and the formalism is quite complicated. But it turns out that in three dimensions, one has a poor man's version of higher spin theories which does not require us to work with the full 2+1 dimensional Vasiliev theory \footnote{The latter in the AdS context is the so-called hs$[\lambda]$ theory, and has been conjectured to be dual to minimal model CFTs in two dimensions \cite{Gop1, Gop2}.}. A simple way to motivate this is to note that 2+1 dimensional gravity can be re-expressed \cite{Carlip} as a Chern-Simons gauge theory with the gauge group $SL(2,\IC)$ \footnote{This specific choice of gauge group corresponds to a positive cosmological constant $\Lambda >0$, which is our main interest in the paper. For the case  $\Lambda =0$, the gauge groups is $ISO(2,1)$ and for  $\Lambda <0$ which is the case sorresponding to AdS$_3$ gravity, its is $SL(2,\IR) \times SL(2,\IR)$.}. It turns out that increasing the gauge group rank from $2$ to $N$ and working with an $SL(N, \IC)$ gauge theory corresponds to working with a spin-2 gravity theory in de Sitter space, coupled to spins ranging from $s=3, ..., N$. We will only be concerned with the $SL(3, \IC)$ theory in this paper, but it is evident that generalizations of our statements exist for any $N \ge 3$.

It is known that 3-D gravity has no propagating degrees of freedom %\footnote{In fact, this statement holds true both in the spin-2 $SL(2,\IC)$ theory as well as in its higher spin cousins.} 
and therefore all its solutions must be locally (a patch of) 2+1 dimensional de Sitter. In particular, quotients of dS$_3$ are also solutions of the theory, analogous to the case of BTZ black holes in the AdS$_3$ case. The quotients of dS$_3$ theory that we will consider are cosmological solutions that contain big bang/big crunch simgularities. Our goal in this paper will be to show that these cosmological singularities can be ``resolved" by a choice of gauge in the Chern-Simons formalism. This is the de Sitter analogue of the observation that higher spin black holes in AdS$_3$ have horizons and singularities that are gauge-dependent \cite{Kraus1, Kraus2, Maloney}. The essential physics behind this is intuitively plausible: the gauge symmetry of spin-2 fields is diffeomorphism invariance (and the implicit freedom associated to local Lorentz rotations of the frames). Diffeomorphisms are a statement about the redundancies of the spacetime decsription, so one might expect that the introduction of higher spin gauge symmetries can bring forth even more dramatic redundencies in the spacetime picture. What we observe is to be understood as a manifestation of this fact: the existence of the cosmological singularity in the metric depends on the choice of the higher spin gauge. It is tempting to speculate that higher spin theories require a generalization of the manifold picture of spacetime \cite{V1, V2}.

In the next section, we give a brief introduction to pure de Sitter gravity as an $SL(2,\IC)$ Chern-Simons theory, and also present $SL(3, \IC)$ gauge theory which is the (spin-3) higher spin theory that we will work with. The goal of this section is partly to fix our notation (see also the Appendices). Section 3 is devoted to the description of the quotient space, which has an interpretation as a cosmology. We describe this geometry first in a form that is analogous to that of the BTZ black hole in AdS$_3$, as well as in a Fefferman-Graham-like form that readily shows the cosmological nature of the spacetime. We also present the solution in the gauge field language and compute its holonomy. Section 4 presents the main point of the paper - we construct a class of higher spin gauge transformations that preserve the holonomy of the solution, but which can desingularize the geometry, and check that indeed the resultant solution has a smooth metric (and higher spin field) everywhere. %Section 5 attempts to make a connection with the Euclidean theory. This section is much more speculative than the rest of the paper, but we believe there is some truth in it, so we will present it here anyway. We make remarks about why the divergence of the Euclidean partition function seen in the de Sitter Farey Tail approach might be tied to the existence of unresolvable singularities in the Lorentzian theory.
The appendices contain some of the technical details.

There has been a lot of work recently on higher spin theories in AdS spaces and their minimal model duals \cite{Gop1, Gop2, Kraus1, Kraus2, Campo, Marc, Justin1, Tomas, JDB}. Not much work has been done in the context of de Sitter, but see \cite{Ouyang} for discussions on higher spin dS/CFT. Our work deals with the bulk geometry exclusively, so the discussions and challenges in that paper regarding the CFT do not concern us. See also \cite{S2, Denef, Das, Shailesh}.

\section{(Higher Spin) Gravity in 2+1 Dimensional de Sitter Space\label{hsds}}

Gravity in three dimensions can be written as a Chern-Simons gauge theory. We will be interested in de Sitter gravity. Raising an lowering the local Lorentz indices using
$\eta_{a b}={\rm diag}\{-1,+1,+1\}$ and using the $SL(2)$ generators\footnote{Note that the basis of $SL(2)$ generators and their algebra are abstract objects. But the nature of the resulting theory would depend on the field in which the coefficients are taken. We will be interested in $SL(2, \IC)$ in this paper.}
$T_a$ that satisfy 
\bea
[T_a, T_b]=\epsilon_{abc}T^c,
\eea
the translation between the gauge field language and the gravity (ie., vielbein and spin connection) language can be written as
\bea
A=\Big(\omega^a_{\mu}+\frac{i}{l}e^{a}_{\mu}\Big)T_a \ dx^\mu, \\
\tilde A=\Big(\omega^a_{\mu}-\frac{i}{l}e^{a}_{\mu}\Big)T_a \ dx^\mu. 
\eea
Here $l$ is the de Sitter length scale and we have defined $\omega_a=\frac{1}{2}\epsilon_{abc}\omega_{\mu}^{bc}dx^\mu$ in terms of the usual spin connection with two tangent space indices\footnote{We use $\epsilon^{012}=1$.}. %{\bf Is this correct?}}. 
This last construction works only in three dimensions, and this is the reason why the Chern-Simons formalism is natural in three dimensions. In terms of these gauge field variables, the Einstein-Hilbert action with cosmological constant takes the form
\bea
S=\frac{k}{4\pi y_R} \int {\rm Tr} \left( A\wedge dA + \frac{2}{3} A\wedge A \wedge A\right) 
-\frac{k}{4\pi y_R} \int {\rm Tr} \left(\tilde{A}\wedge d\tilde{A} +\frac{2}{3} \tilde{A}\wedge \tilde{A}\wedge \tilde{A}\right).
\label{EHCS}
\eea
where $y_R$ is defined via ${\rm Tr}(T_a T_b)=\frac{y_R}{2}\eta_{ab}$. We will work with $y_R=4$ in this paper.  
To make the action real, we need to choose\footnote{This identification of the two gauge fields is one of the technical reasons why the dS$_3$ results are not merely a trivial ``$l$ replaced with $il$" version of the AdS$_3$ results.} $\tilde A_a\ T^a=A^{*}_a\ T^a$, where it is important that the $T^a$'s don't get conjugated. Then, the flatness conditions $F=0=\tilde F$ turns into the Einstein equations when we choose the Chern-Simons level to be
\bea
k=-\frac{i l}{4 G}.
\eea
%{\bf The represnetation correction.} 
We will set $8G=1$ in what follows. This is the connection between 3D de Sitter gravity and Chern-Simons gauge theory. 

As mentioned in the introduction, working with higher spins is a complicated business in dimensions higher than 3. But in three dimensions, the Chern-Simons language allows a simple way to write down interacting higher spin theories, by increasing the rank of the Chern-Simons gauge group. We will be dealing with positive cosmological constant in this paper, and for simplicity we will restrict ourselves to $SL(3,\IC)$ Chern-Simons gauge theory. This corresponds to a spin-3 field coupled to gravity in de Sitter space. Explicitly, we introduce the extra generators $T_{ab}$ to the $T_a$ of $SL(2)$ and the full $SL(3)$ algebra takes the form
\bea
&&[T_a,T_b] = \epsilon_{abc} T^c, \\
&&[T_a,T_{bc}] = \epsilon^d_{\ \ a(b} T_{c)d}, \\ 
&&[T_{ab},T_{cd}] = \sigma \left(\eta_{a(c} \epsilon_{d)be}+\eta_{b(c} \epsilon_{d)ae}\right)T^e.
\eea
The $T_{ab}$ are symmetric and traceless and therefore are five in number, adding up to a total of eight generators for $SL(3)$, as expected. Its clear from the algebra that the constant $\sigma$ can be gotten rid of by absorbing it into the $T_{ab}$ generators - it will not affect the content of our discussion, so we will choose it to be -1, in parallel with the AdS$_3$ case discussed in the literature \cite{Maloney, Kraus1}. %{\bf How does this affect the fact that the gauge group here is complexified? Reconcile the various reality properties with the discussion at the end of section 2.2 in Ouyang.} 

The above embedding of the $SL(2)$ algebra generators in $SL(3)$ is called principal embedding, and this is what we will be using in this paper.

With this enlarged gauge group, one now considers a Chern-Simons theory with the gauge field defined by
\bea
A = \left(\omega_{\mu}^{\;\; a} + \frac{i}{\ell} e_{\mu}^{\;\; a} \right) T_a dx^{\mu}+
\left(\omega_{\mu}^{\;\; ab} + \frac{i}{\ell} e_{\mu}^{\;\; ab} \right) T_{ab} dx^{\mu},
\eea
and its complex conjugate $\tilde A$, and then looks at the same action (\ref{EHCS}) as before. This theory is a theory of gravity coupled to a spin-3 field. By simple index counting, the obvious candidates for the metric and the spin-3 field (appropriately normalized) are \cite{Campo}
\bea
g_{\mu\nu}=\frac{1}{2}{\rm Tr}(e_{(\mu}e_{\nu)}), \ \ \psi_{\mu\nu\alpha}=\frac{1}{9}{\rm Tr}(e_{(\mu}e_{\nu}e_{\alpha)}), \ \ {\rm with} \ \ e_{\mu} \equiv e_{\mu}^aT_a+ e_{\mu}^{\;\; ab} T_{ab}.
\eea
When perturbed around the dS$_3$ background, the $e_{\mu}^{\;\; ab}$ satisfy the Fronsdal equations of motion \cite{Blencowe}, %({\bf CHECK if the paper explicitly talks about de Sitter case.)}, 
and therefore justify their identification as a spin-3 field. Just as diffeomorphisms and local Lorentz invariance of the spin-2 theory are identified with the $SL(2)$ part of the algebra, the generators $T_{ab}$ correspond to the ``higher spin gauge symmetry".

We find it is convenient to relate the above form of the algebra to the $L_m, W_n$ generators defined by \cite{Kraus1, Maloney}\footnote{An explicit form of these $L_m, W_n$ generators as well as the algebra they satisfy is given in an Appendix.} so that we can adapt their results and notations. In particular, we choose
\bea
T_{0}=-iL_{0},\qquad T_{1}=i\left(\frac{2L_{1}+L_{-1}}{2\sqrt{2}}\right),\qquad T_{2}=\frac{2L_{1}-L_{-1}}{2\sqrt{2}}.\label{TintermsofL} 
\eea
From the definition of the $L_m$'s in the Appendix, it is straightforward to check that the resulting $T_a$ satisfy the $SL(2)$ algebra\footnote{The choice of $T_a$ in terms of $L_m$ considered in \cite{Ouyang} does not reproduce the $SL(2)$ algebra: we believe this is a typo, because the rest of the claims about the bulk theory there seem reasonable and correct.}. %{\bf Check the rest of the traces in Avinash and Shubho notebooks.}

%{\bf Write down the map relating the $W_n$'s and the $T_{ab}$ in the Appendix. Maloney writes them down. Are we using the same ones? Is there any place where the explicit form of the relation between the $W_n$'s and the $T_{ab}$ matter for us? If so, do we want to use the Ouyang defintion? Has he bungled that too? Does the choice of $\sigma$ or the difference between $\IR$ vs $\IC$ matter?}

%{\bf DISCUSS GAUGE TTRANFORMATIONS AND THE BOUNDARY CONDITIONS AT FUTURE INFINITY.}

The fact that (higher spin) de Sitter gravity has a Chern-Simons formulation means that the solutions are invariantly characterized by the holonomies of the gauge field. This is a fact that we will use in the later sections.

\section{The Quotient Cosmology and its Holonomy\label{ppc}}

Solutions of 3-dimensional de Sitter gravity are locally dS$_3$. So analogous to the BTZ black hole which can be thought of as a quotient of AdS$_3$, the solutions of dS$_3$ gravity can be thought of as quotients of dS$_3$. In this section, we will discuss such a quotient \cite{Park, Bala} which is a time-dependent background with cosmological simgularities. We will compute its holonomy. In the next section, we will show that (higher spin) gauge transformations that preserve the holonomy can change the metric so drastically that the singularity is gone in the final metric.

General quotient spacetimes of dS$_{3}$
were constructed in \cite{Bala}. A subclass of these, which they call
Kerr-dS$_{3}$ are regular quotients - they are the dS counterparts
of BTZ black holes in AdS and have a topology $\IR^{2}\times S^{1}$.
However, another subclass of quotients exists which have a topology
of $\IR\times S^{1}\times S^{1}$. These are singular quotients
with big bang/big crunch like singularities and a single spacelike
asymptotic region (infinite future/past), locally resembling the geometry
outside the cosmological horizon of Kerr-dS$_{3}$. This geometry
is more appropriately thought of as a cosmology, so we will call it
a quotient cosmology. We will describe it in the next subsection, but we start with the (closely related) Kerr-dS$_{3}$ metric in BTZ like form:
%The de Sitter quotient metric that we are interested in is written down in \cite{Bala}. There, they call it the Kerr-dS$_3$ geometry and write it in a form closely parallel to the BTZ black hole. But the region that contains the asymptotic region in this geometry is more appropriately thought of as a cosmology, so we will call it a quotient cosmology. The metric in BTZ like form takes the form:
\[
ds^{2}=-N^{2}(r)dt^{2}+N^{-2}(r)dr^{2}+r^{2}\left(N_{\phi}dt+d\phi\right)^{2}
\]
with
\begin{equation}
N^{2}(r)=M-\frac{r^{2}}{l^{2}}+\frac{J^{2}}{4r^{2}},\qquad N_{\phi}=-\frac{J}{2r^{2}},\label{eq: Kerr-dS_3 inside}
\end{equation}
%{\bf Discuss some properties of the metric, like the big-bang big-crunch structure, etc.}

One can check that the following $SL(2,\IC)$ Chern-Simons connection can reproduce this metric:
\begin{equation}
A^{0}=N(r)\left(d\phi+i\frac{dt}{l}\right),\quad A^{1}=\frac{l\, N_{\phi}-i}{N(r)}\frac{dr}{l},\quad A^{2}=\left(rN_{\phi}+i\frac{r}{l}\right)\left(d\phi+i\frac{dt}{l}\right).\label{eq: SL(2,C) connection for axisymmetric}
\end{equation}
% This is easy to guess by adapting the corresponding expressions of \cite{Lowe:2008ye,Cangemi:1992my} by changing the sign of the cosmological constant i.e. making the AdS radius imaginary, \[ l\rightarrow il.\]
% Now one can substitute,  \begin{equation} N^{2}(r)=M-\frac{r^{2}}{l^{2}}+\frac{J^{2}}{4r^{2}},\qquad N_{\phi}=-\frac{J}{2r^{2}},\label{eq: Kerr-dS_3 inside} \end{equation}
% to obtain the corresponding gauge field for the Kerr-dS$_{3}$ metric of \cite{Balasubramanian:2001nb}. The gauge connection for this metric is 
%\[ A^{a}=A_{\mu}^{a}dx^{\mu}=\omega_{\mu}^{a}+i\frac{e_{\mu}^{a}}{l}. \]
% and  \[ \left[T_{a},T_{b}\right]=\epsilon_{ab}\,^{c}T_{c},\quad\epsilon^{012}=1. \]
In the context of higher spin SL(3) gravity, we use the set of $SL(3)$
generators provided in \cite{Kraus1} to embed the $SL(2)$, the specific choice we make is given in (\ref{TintermsofL}). The gauge field will then be given by $A=A^a T_a$. %{\bf Are we making sure in all this, that $T^a=\eta^{ab}T_b$ ?? The local Lorentz metric, it is not $\delta_{ab}$, because ${\rm Tr}(T_aT_b)\sim \eta_{ab}$, not $\delta_{ab}$. I am worried that you guys might have forgotten this - the metric on the algebra is supposed to be the local Lorentz metric, if $A=A^aT_a$ should equal $A_aT^a$.}

Our first goal is to compute the holonomy of this connection. As done in \cite{Shubho}
we need to solve for $U\in$ SL$(2,\IC)$ such that 
\[
A=U^{-1}dU.
\]
These are the solutions to the flatness condition $F=0$. %({\bf Are these the most general solutions?  Sure, we allow for multi-valuedness, but how does one show that these are the only solutions even after that (up to gauge transformations)?}). 
Of course, the $U$'s that we get are not necessarily single valued, and thats where the holonomy information is captured. 
As is worked out in the Appendix, for the quotient cosmology gauge field above,
the solution is, 
\[
U=e^{\theta_{0}T_{0}}e^{\theta_{1}T_{1}},
\]
 for, 
\begin{equation}
\theta_{0}=\sqrt{M+i\: J/l}\left(\phi+it/l\right),\label{eq: theta0}
\end{equation}
\begin{equation}
\cosh\theta_{1}=\frac{N(r)}{\sqrt{M+iJ/l}},\qquad\sinh\theta_{1}=-\frac{rN_{\phi}(r)+ir/l}{\sqrt{M+iJ/l}}.\label{eq:theta1}
\end{equation}
From $U$, one can extract the Wilson loops for loops
enclosing $r=0$ (at constant time in the $\phi$--direction), 
\begin{equation}
W(A)=\exp(\oint A)=U^{-1}(t,r,\phi=0)U(t,r,\phi=2\pi)\label{eq: Wilson loop}
\end{equation}
 and further, the eigenvalues of the holonomy matrix, $w$ defined
by $W=\exp(w)$, by exponentiating the eigenvalues of the Wilson loop
$W$.

%\begin{equation} 
%W(A)=\left(\begin{array}{ccc}
%\left(\cosh^{2}\frac{\theta_{1}}{2}e^{-i\theta_{0}}-e^{i\theta_{0}}\sinh^{2}\frac{\theta_{1}}{2}\right)^{2} & \frac{i}{\sqrt{2}}\left(\left(1-\cos\theta_{0}\right)\cosh\theta_{1}+i\sin\theta_{0}\right)\sinh\theta_{1} & \sin^{2}\frac{\theta_{0}}{2}\sinh^{2}\theta_{1}\\

%\frac{i}{\sqrt{2}}\left(\left(1-\cos\theta_{0}\right)\cosh\theta_{1}+i\sin\theta_{0}\right)\sinh\theta_{1} & \cosh^{2}\theta_{1}-\cos\theta_{0}\sinh^{2}\theta_{1} & -\frac{i}{\sqrt{2}}\left(\left(1-\cos\theta_{0}\right)\cosh\theta_{1}-i\sin\theta_{0}\right)\sinh\theta_{1}\\

%\sin^{2}\frac{\theta_{0}}{2}\sinh^{2}\theta_{1} & -\frac{i}{\sqrt{2}}\left(\left(1-\cos\theta_{0}\right)\cosh\theta_{1}-i\sin\theta_{0}\right)\sinh\theta_{1} & e^{i\theta_{0}}\left(\cosh^{2}\frac{\theta_{1}}{2}-e^{-i\theta_{0}}\sinh^{2}\frac{\theta_{1}}{2}\right)^{2}

%\end{array}\right)\label{eq: Wilson loop for Kerr-dS_3}
%\end{equation}
 
The Wilson loop is straightforward to compute, and its eigenvalues are given by $e^{\lambda}$ with
\bea
\lambda=0, \pm\left(2\pi i\sqrt{M+i\: J/l}\right).
\eea 
The holonomy matrix turns out to be, 
\begin{equation}
w=\left(\begin{array}{ccc}
0\\
 & -2\pi i\sqrt{M+i\: J/l}\\
 &  & 2\pi i\sqrt{M+i\: J/l}
\end{array}\right).\label{eq: Holonomy matrix for Kerr-Ds_3 around singualrity}
\end{equation}

%{\bf Does this compare with the Kraus-Gutperle holonomy? There it seemed that the holonomy was just $2 \pi i$ etc for BTZ? What am I forgetting? } 

\subsection{Fefferman-Graham in de Sitter}

The metric in the form (\ref{eq: Kerr-dS_3 inside})
is best suited for $r<r_{+},$ where 
\[
r_{\pm}^{2}=l^{2}\left(\sqrt{M^{2}+\left(J/l\right)^{2}}\pm M\right)/2.
\]
 In order to conduct an asymptotic symmetry analysis one needs continue the
metric beyond the horizon. Now $r$ becomes a time coordinate.
To facilitate comparison with the prevailing higher spin literature
\cite{Kraus1, Campo} lets switch to new coordinates:
%\[
$w=\phi+it/l,\ \ \bar{w}=\phi-it/l$.
%\]
We will also introduce a new time coordinate, $\tau$, to replace the (now time-like) $r$ coordinate:
%it can be checked that the coordinate transformation one needs to bring the metric from the BTZ-like form to the Fefferman-Graham form is
\[
\tau=\ln\left(\frac{\sqrt{r^{2}-r_{+}^{2}}+\sqrt{r^{2}+r_{-}^{2}}}{2l}\right)
\]
%(Perhaps as is done in \cite{Balasubramanian:2001nb}we would be better off swapping $\rho$ and $t$). 
This new coordinate system expresses the quotient cosmology in a de Sitter version of the Fefferman-Graham coordinates. From now on, since the $t$ coordinate is no longer a time coordinate, we will rename it as $z$. Therefore, really, $w$ and $\bar w$ are defined as
\[
w=\phi+iz/l,\qquad\bar{w}=\phi-iz/l
\]
In the quotient cosmology metric of \cite{Bala} which has big bang/crunch singularities, one identifies\footnote{The periodicity of the solution is arbitrary. We choose it by an analogy with the Euclidean BTZ metric, but there is nothing sacred about that choice for our purposes.} $z \sim z+(2 \pi l^2  r_+)/ (r_+^2 +r_-^2)$, so henceforth this will be implicit in our discussion. So the topology is now $\IR \times S^1 \times S^1$ instead of the $\IR^2 \times S^1$ of Kerr-dS$_3$.

We can generalize this more. In a Fefferman-Graham gauge, the most general asymptoically dS$_3$ solution to pure gravity with a positive cosmological constant in $2+1$ D can be written down. This is the dS$_3$ analog of the result  by Banados \cite{Banados:1998gg} for AdS$_3$.
%\footnote{Can someone check this, I've only checked it for 
 % 
Such a general asymptotically dS$_3$ metric can be written in terms of one complex function $L(w)$ and its complex conjugate\footnote{We thank Avinash Raju for checking that this metric satisfies Einstein equation with a positive cosmological constant.}:
\begin{equation}
ds^{2}=\frac{l}{2}\left(L(w)dw^{2}+\bar{L}(\bar{w})d\bar{w}^{2}\right)+\left(l^{2}e^{2\tau}+\frac{L(w)\bar{L}(\bar{w})}{4}e^{-2\tau}\right)dwd\bar{w}-l^{2}d\tau^{2}.\label{eq: SL(2) sector general solution}
\end{equation}
The quotient cosmology of the previous subsection corresponds to the case $L$, $\bar{L}$ constant.

%{\bf Avinash's check treats $w$ and $\bar w$ as real coordinates. I don't know if RGTC's internal wiring relies on them being real. I doubt it, but its probably worthwhile to make one check:  The cosmological constant that he finds is $2/l^2$. In our sign conventions, we need to make sure that the sign of that matches that of dS$_3$, not of AdS$_3$! If we are finding a positive cosmological constant, then yes, we have checked it definitively. I just want to make sure that we are not secretly just checking the AdS case again. Its unlikely, because the $d\rho^2$ in the metric has the wrong sign, but lets be sure. So we need to check that the sign of the cosmological constant is also the correct one!}

 The asymptotic symmetry analysis of \cite{Bala}
identifies 
\begin{equation}
L+\bar{L}=Ml,\quad L-\bar{L}=iJ.\label{eq: dS_3 asymptotic charges}
\end{equation}
The comparison of the asymptotic analysis between AdS$_3$ and dS$_3$ is straightforward \cite{Ouyang}. One can go back and forth between AdS$_{3}$ and dS$_{3}$ by using the
following short cut identifications/replacements, 
\[
l_{AdS}\rightarrow il_{dS}
\]
 
\[
L_{AdS},\bar{L}_{AdS}\rightarrow-iL_{dS},-i\bar{L}_{dS}
\]
 
\[
M_{AdS}\rightarrow-M_{dS}
\]
 
\[
J_{AdS}\rightarrow J_{dS}
\]
 For example the usual asymptotic relations in AdS$_{3}$ are: 
\[
L+\bar{L}=Ml,\qquad L-\bar{L}=J
\]
 can be used to arrive at the dS$_{3}$ asymptotic relations of \cite{Bala}.
\[
L+\bar{L}=Ml,\qquad L-\bar{L}=iJ.
\]
Note however that this quick-fix match between AdS$_3$ and dS$_3$ is of limited use and does not exist in many other contexts. For example, the gauge transformations and the final form of the metric that we will discuss are entirely different from their AdS$_3$ analogues. 

A convenient choice of frame fields for this general FG metric is\footnote{We are suppressing
the $w$ / $\bar{w}$ dependence of $L$ / $\bar{L}$}:
\begin{eqnarray*}
e^{0} & = & ld\tau\\
e^{1} & = & -\frac{i\: l}{2}\left(e^{\tau}-\frac{L}{2l}e^{-\tau}\right)dw+c.c.\\
e^{2} & = & \frac{l}{2}\left(e^{\tau}+\frac{L}{2l}e^{-\tau}\right)dw+c.c.
\end{eqnarray*}
 The corresponding $SL(2,\mathbb{C})$ connection is, 
\begin{equation}
A=i\, T_{0}\, d\tau+\left[\left(e^{\tau}-\frac{L}{2l}e^{-\tau}\right)\, T_{1}+i\,\left(e^{\tau}+\frac{L}{2l}e^{-\tau}\right)T_{2}\right]dw\label{eq: SL(2,C) connection for general metric}
\end{equation}
and its conjugate. %{\bf  Note that we have {\em not} chosen $L=\bar{L}$.}

We also need the $SL(2,\mathbb{C})$ group element generating this.
To this end we first rewrite the connection in the form, 
\[
A^{0}=d\psi_{0},\qquad A^{1}=-\sin\psi_{0}\: d\psi_{2},\qquad A^{2}=\cos\psi_{0}\: d\psi_{2}
\]
 with, 
\[
\psi_{0}=i\left(\tau-\frac{1}{2}\ln\frac{L}{2l}\right),\qquad\psi_{2}=i\sqrt{\frac{2L}{l}}\omega.
\]
 Now it can be shown that one can write (see Appendix), $A=V^{-1}dV$
for 
\[
V=e^{\psi_{2}T_{2}}e^{\psi_{0}T_{0}}.
\]
 which in turn gives the holonomy matrix, $w$, 
\[
w=\left(\begin{array}{ccc}
0\\
 & -2\pi\, i\,\sqrt{\frac{2L}{l}}\\
 &  & 2\pi\, i\,\sqrt{\frac{2L}{l}}
\end{array}\right)
\]
 which is identical to (\ref{eq: Holonomy matrix for Kerr-Ds_3 around singualrity})
once we substitute, $L=\left(Ml+iJ\right)/2.$ (Holonomies are diffeomorphism
invariant).
%for the $SL(2,\mathbb{C})$ metric (we can always do that for asymptotically dS boundary conditions including the quotient cosmology%dS$_{3}$ quotients

For future reference, we define Fefferman-Graham gauge to be 
\[
g_{\tau\tau}=-l^{2},\qquad g_{\tau w}=g_{\tau\bar{w}}=0.
\]
 and vanishing spin-3 field. In terms of $SL(3)$ gauge theory, one way we can realize this gauge is by turning on just the principally embedded $SL(2)$ sector
in the particular form, 
\[
A_{\tau}=\pm i\, lT_{0},\quad A_{w}=A_{w}^{a}(\tau, w)T_{a},\quad A_{\bar{w}}=0.
\]
%{\bf (Have we explicitly checked that this gauge field gives rise to the above conditions on the FG gauge metric?)} 
The connection (\ref{eq: SL(2,C) connection for general metric}) corresponding to the general Banados type solution satisfies this condition. 
%By analogy with \cite{Campo} for AdS$_3$ we define asymptotic dS condition as \cite{Ouyang} \[ \lim_{\tau\rightarrow\infty}A-A_{dS}\sim O(1).\]
 % 

\section{Resolution of the Cosmological Singularity}

The quotient cosmology metric, where we set $L$ and $\bar L$ to constant, takes the following explicit form in Fefferman-Graham gauge (note: we have renamed $t$ to $z$): 
\[
ds^{2}=-l^{2}d\tau^{2}+\left|e^{\tau}-\frac{L}{2l}e^{-\tau}\right|^{2}dz^{2}+\left|le^{\tau}+\frac{L}{2}e^{-\tau}\right|^{2}d\phi^{2}+i\,\frac{L-\bar{L}}{2}dz\, d\phi.
\]
At this stage, we will restrict our attention to the case $L=\bar{L}$ for simplicity. It is easy to see that the metric component $g_{zz}$ vanishes at $\tau=\frac{1}{2}\ln\frac{L}{2l}$ in this case. In the original Schwarzschild coordinates this corresponds to the vanishing of $N(r)^2$ and was identified as big bang/big crunch singularities in causal structure \cite{Bala}, due to the periodic identification in $z$. The segment $-r_+ < r < r_+$ has to be excised: extending the metric beyond these points results in closed time like curves.

%It turns out that one has Big-Bang/big crunch like singularities at $\tau=\frac{1}{2}\ln\frac{|L|}{2l}$ although the metric looks totally regular. Although not manifest inFefferman-Graham coordinates, the big bang/big crunch singularities are apparent in the BTZ/Schwarzschild coordinates $t,r,\phi$ at $r=r_{+}$ (with periodic spacelike $t;\phi$) as was pointed out in \cite{Bala}. Singularities arise at the point where $N(r)^{2}$ vanishes, verging on the (excised) region containing closed timelike curves. However,if $L=\bar{L}$, i.e. the non-rotating case, this singularity becomes apparent even in the Fefferman-Graham coordinates as a metric singularity where $g_{zz}$ vanishes at $\tau=\frac{1}{2}\ln\frac{L}{2l}$. 
In what follows, we will conduct a SL(3) gauge transformation on the
non-rotating quotient cosmology. This will result in a positive definite
$g_{zz}$ for all $\tau$, thus eradicating the singularity. Note that the singularity we will be resolving is not a curvature singularity: the curvature is finite and constant everywhere else as it should be for a constant curvature space. This is the reason why we call this a ``toy" big bang.

%(By the same token this same trick will also work for even for the $L\neq\bar{L}$ case, resulting in a shift of $N(r)^{2}$ making it positive semi-definite for all $t$ case, although that will not be visible in the Fefferman-Graham coordinates).

%It has vanishing $g_{zz}$ at $\tau=\frac{1}{2}\ln\frac{L}{2l}$.  Equivalently, it was pointed out in \cite{Bala} the metric in $t,r,\phi$ coordinates has big bang/big crunch singularities at $r=\pm r_{+}$ with periodic spacelike $t,\phi$. Note that this is a causal structure singularity and not a curvature singularity: the curvature is finite and constant everywhere else as it should be for a constant curvature space.

Before proceeding further, we make one observation. Consider the connection,
\begin{equation}
a_0=\left[\left(1-\frac{L}{2l}\right)T_{1}+i\left(1+\frac{L}{2l}\right)T_{2}\right]dw.\label{eq: Kerr-dS3 curly}
\end{equation}
%for some as yet arbitrary $\alpha$. One can obtain the Kerr-dS$_{3}$
The quotient cosmology connection of Eq. (\ref{eq: SL(2,C) connection for general metric})
can be obtained by performing a \textbf{single valued} gauge transformation on this primitive connection $a_0$:
\bea
A=b^{-1}a_0\:b+b^{-1}db, \label{firsta}
\eea
for 
\[
b(\tau)=\exp\left(i\tau T_{0}\right)=\exp(\tau L_{0})
\]
This is because $b$ is a sole function of $\tau$ and is therefore single
valued in the $\phi$ direction.

\subsection{The class of singularity resolving spin 3 gauge transformations\label{sec:General Case}}
 
The aim of this subsection is to obtain a fairly general set of holonomy
preserving gauge transformations which take our quotient cosmology  %$SL(2,\mathbb{C})$
metric with singularities to a regular $SL(3,\mathbb{C})$ metric.
The understanding is that making metric components non-vanishing eliminates
all metric singularities which may or may not be true singularities.

%Our strategy is to first start with a class of gauge fields that contains (\ref{eq: Kerr-dS3 curly})  as a special case. Then we demand

The algorithm is as follows: 
\begin{enumerate}
\item We first define a connection $a \in SL(3)$ which is a generalization of the $a_0$ defined in the previous subsection: 
\bea
a=a_0+Y\, dw=\left[\left(1-\frac{L}{2l}\right)T_{1}+i\left(1+\frac{L}{2l}\right)T_{2}\right]dw+Y\, dw,\qquad Y=\sum_{a=-2}^{2}C_{a}(\tau)W_{a}.\label{seconda}
\eea
%again in the same gauge. 
The explicit matrix realizations
of the SL(3) generators, $W_{a}$ are provided in Appendix C.
Flatness of this connection i.e. $F=da+a\wedge a=0$
demands the coefficients $C_{a}$'s have no $\tau$ dependence i.e.
they be constants (See Appendix \ref{sub:Flatness condition}). Note that the $C_a$'s are five complex numbers, so this is 10 parameters worth of freedom.

\item We will apply the gauge transformation, $U(\tau)=\exp(i\tau T_{0})=\exp(\tau L_{0})$
on $a$ and obtain connection, $A'$:
\begin{equation}
A'=A+Xdw,\label{eq: regular SL(3)}
\end{equation} 
where
\[
X\equiv\exp(-L_{0}\tau)Y\exp(L_{0}\tau)=\sum_{a=-2}^{2}e^{a\tau}C_{a}W_{a}.
\]
Note that here $A$ is the connection that gives rise to the quotient cosmology metric. Our goal will be to look for gauge fields $A'$ in this class that have the same holonomy as the quotient cosmology connection $A$, but which give rise to non-singular metrics. 
%We will consider the $SL(3)$ metric derived from $A'$.

\item We will demand that the $g_{zz}$ and $g_{\phi\phi}$ arising from $A'$
differ by a positive quantity from those arising from $A$ (while not affecting the rest of the
metric components). This is a sufficient condition for metric regularity. This requires%
\footnote{This is easy to demonstrate. The new vierbein derived from $A'$ is,
\[
e'=e+\frac{Xdw-\bar{X}d\bar{w}}{2i/l}.
\]
Then the new metric components are:
\[
g'_{ww}=g_{ww}-\frac{l^{2}}{8}Tr(X^{2}),\quad g'_{\bar{w}\bar{w}}=g_{ww}-\frac{l^{2}}{8}Tr(\bar{X}^{2}),\quad g'_{w\bar{w}}=g_{w\bar{w}}+\frac{l^{2}}{8}Tr\left(X\bar{X}\right).
\]
 Noting that,
\[
2g'_{w\bar{w}}dwd\bar{w}=2g'_{w\bar{w}}\left(\frac{dz^{2}}{l^{2}}+d\phi^{2}\right),
\]
if we arrange so that, $Tr(X^{2})=Tr(\bar{X}^{2})=0$ and $Tr(X\bar{X})>0$
we add positive definite numbers to $g_{zz}$ and $g_{\phi\phi}$
removing any zeros present in them.%
}, 
\[
{\rm Tr}(X^{2})={\rm Tr}(\bar{X}^{2})=0,\qquad {\rm Tr}(X\bar{X})>0
\]
 which are 2 (real) equations and one constraint inequality: 
\begin{equation}
C_{0}^{2}-3C_{1}C_{-1}+12C_{2}C_{-2}=0,\label{eq: Trace vanish}
\end{equation}
 
\begin{equation}
C_{0}\bar{C}_{0}-3C_{1}\bar{C}_{-1}+12C_{2}\bar{C}_{-2}+c.c>0.\label{eq: metric positive}
\end{equation}
Notice that the equations are $\tau$ independent, even though $X$ is $\tau$-dependent.
\item Finally, we want to ensure that both $A'$ and $A$ have the same holonomy, so that they are related by a single valued gauge transformation. Computing the $U$ matrix and explicitly evaluating the eigenvalues of the holonomy matrix like we did before is tiresome now because there are five more generators coming from the spin 3 charges.  

 Instead we will follow \cite{Maloney} and fix the holonomy matrix in terms of its characteristic polynomials. The idea is that any $SL(3,\IC)$ matrix $M$ 
%for the equation \beaM^3=\Theta_{0} I +\Theta_{1}X+\Theta_{2}X^2, \eea
satisfies the equation $M^3=\Theta_{0} I +\Theta_{1}M$ where
\bea
\Theta_{0}={\rm det} (M), \ \ {\rm and} \ \ \Theta_{1}=\frac{1}{2}{\rm Tr} (M^2)
\eea
So we will equate the characteristic coefficients of the holonomy matrix for $A'=A+Xdw$ with that of $A$. The holonomy matrix is easily computed by integrating the gauge field around the $\phi$-circle 
at fixed $z$. Demanding that the $\Theta_{0}$ and $\Theta_{1}$ are the same for $A$ and $A'$ holonomies %gives:
%s simply defined as the  
%the holonomy of the original $SL(2)$ singular metric, 
gives %2 new(real) constraints(instead of 4 constraints as the $\Theta_{1}$ constraint is the same the as the previous $Tr(X^{2})=0$) 
\begin{equation}
\Theta_{0}:C_{0}^{3}-\frac{9}{4}\frac{L}{l}C_{0}+\frac{27}{2}\left(C_{1}^{2}C_{-2}+C_{-1}^{2}C_{2}\right)-\frac{9}{2}C_{0}\left(C_{1}C_{-1}+8C_{2}C_{-2}\right)+27C_{-2}+\frac{27}{16}\frac{L}{l}C_{2}=0.\label{eq: Theta_zero}
\end{equation}
\begin{equation}
\Theta_{1}:C_{0}^{2}-3C_{-1}C_{1}+12C_{-2}C_{2}=0.\label{eq: Theta_one}
\end{equation}
The first %$\Theta_{0}$ 
equation is obtained straight from\footnote{The specific form of the $\Theta_0$ equation here depends on the specific choice of $T_a$ that we made (\ref{TintermsofL}). There are other choices of $T_a$ in terms of $L_a$ which result in the same ${\rm Tr}(T_a T_b)$ and $SL(2)$ algebra (and therefore the metric), but which can change the determinant that we are computing here. Note that the $\Theta_1$ equation will not change because the traces are protected.}
\[
{\rm Det} \Big(\int d\phi (A+X dw)|_{z={\rm const.}}\Big)-{\rm Det} \Big(\int d\phi (A)|_{z={\rm const.}}\Big)=0
\]%And a similar one for $\Theta_{1}$. Correct?)} 
A similar equation for the ${\rm Tr}$ gives rise to the second equation.

Together these give rise to 2 new real constraints (instead of 4, because the $\Theta_{1}$ constraint is the same as the already found $Tr(X^{2})=0$ constraint).

\item We can simplify the condition (\ref{eq: Theta_zero}) a bit by inserting
(\ref{eq: Trace vanish}) ,
\begin{equation}
C_{0}\left(C_{1}C_{-1}+32C_{2}C_{-2}+\frac{3L}{2l}\right)-3\left(C_{1}^{2}C_{-2}+C_{-1}^{2}C_{2}+6C_{-2}+\frac{3L}{8l}C_{2}\right)=0\label{eq:Theta_zero1}
\end{equation}

\item So we are left with $10-2-2=6$ parameter family of singularity eliminating
spin-3 transformations. The general form of the resultant regular
metric is,
\[
ds^{2}=-l^{2}d\tau^{2}+\left(\left|e^{\tau}-\frac{L}{2l}e^{-\tau}\right|^{2}+\alpha\right)dz^{2}+\left(\left|le^{\tau}+\frac{L}{2}e^{-\tau}\right|^{2}+\alpha l^{2}\right)d\phi^{2}+i\,\frac{L-\bar{L}}{2}dz\, d\phi
\]
with 
\[
\alpha=\frac{2}{3}\left(C_{0}\bar{C}_{0}-3C_{1}\bar{C}_{-1}+12C_{2}\bar{C}_{-2}+c.c\right).
\]

\item A convenient choice is to set, $C_{\pm2}=0$. %{\bf (The gauge tansformations that can accomplish this are not continuously connected to the identity.)} 
Then the equations are:
\[
C_{0}^{2}-3C_{1}C_{-1}=0
\]
\[
C_{0}^{2}-\frac{9}{4}\frac{L}{l}-\frac{9}{2}C_{1}C_{-1}=0
\]
which solves to give,
\[
C_{0}=i\:3\sqrt{\frac{L}{2l}},\quad C_{1}C_{-1}=-\frac{3}{2}\frac{L}{l}
\]
and we also need to make $g_{zz},g_{\phi\phi}$ positive definite,
i.e. satisfy 
\[
\frac{9}{2}\frac{L}{l}-3C_{1}\bar{C}_{-1}+c.c>0
\]

\item As an example, for the case when asymptotic charge $L>0$, we can
further choose, $C_{\pm1}\in\mathbb{R}$ %{\bf (Why is this important?)} 
and then we have automatically
satisfied the metric positivity constraint,
\[
|C_{0}|^{2}-3C_{1}\bar{C}_{-1}+c.c=18\frac{L}{l}>0
\]
and the connection, 
\bea
a=\left[\left(1-\frac{L}{2l}\right)T_{1}+i\left(1+\frac{L}{2l}\right)T_{2}\right]dw+\left(C_{1}W_{1}+C_{-1}W_{-1}+i\:3\sqrt{\frac{L}{2l}}W_{0}\right)dw, \label{finalgauge}
\eea
and the metric,
\begin{equation}
ds^{2}=-l^{2}d\tau^{2}+\left(\left|e^{\tau}-\frac{L}{2l}e^{-\tau}\right|^{2}+\frac{6L}{l}\right)dz^{2}+\left(\left|e^{\tau}+\frac{L}{2l}e^{-\tau}\right|^{2}+\frac{6L}{l}\right)l^{2}d\phi^{2}.\label{eq:Regular metric}
\end{equation}
%{\bf (There is a contradiction between $\frac{9L}{2l}$ and the expression for $\alpha$ above. The expression for $\alpha$ seems fishy.)}
One can easily check that at the erstwhile singularity, $\tau=\frac{1}{2}\ln\frac{L}{2l}$
the scalar curvature is currently finite, $R=\frac{176}{117}\frac{1}{l^{2}}$. %{\bf CHECK!!!!} 
We have checked that the curvature scalars are finite for all finite values of $\tau$. 

%{\bf (Add the expression for Ricci and Riemann$^2$ so that it is clear that there are no singularities anywhere.)}

\item To appreciate that we have regularized the big-bang/big-crunch
we revert to Schwarzschild coordinates which cover regions beyond
$r_{+}$ using the coordinate transformation (unnumbered equation,
the first equation on page 7). In Schwarzschild coordinates the quotient
cosmology metric after the spin-3 transformation is,
\[
ds^{2}=-\frac{l^{2}}{(r^{2}-r_{+}^{2})}dr^{2}+\left(\frac{r^{2}-r_{+}^{2}}{l^{2}}+\alpha\right)dz^{2}+\left(r^{2}+\alpha l^{2}\right)d\phi^{2}.
\]
Earlier for the quotient cosmology metric the region, -$r_{+}<r<r_{+}$
had to be excised as it contained closed timelike curves ($g_{zz}$
turns negative with $z$-direction having a topology of $S^{1}$).
Due to this excision we were left with big-bang (and big-crunch) singularity \cite{Bala} 
at $r_{+}$ (and $-r_{+}$). Now however the addition of $\alpha=6L/l$
makes $g_{zz}$ positive definite in the whole interval ($g_{zz, {\rm min}}=g_{zz}(r=0)=\alpha-r_{+}^{2}/l^{2}=4\frac{L}{l}>0$)
and everywhere else! This explicitly shows that we have removed the (causal structure) singularities that were originally present in the metric.
\end{enumerate}

We make one observation. The specific gauge field configuration that we have chosen (\ref{finalgauge}) is not continuously connected in the $C_a$ parameter space with the original connection $A$ corresponding to the quotient cosmology. This is because we have imposed the condition $C_1C_{-1}={\rm const}$, so $C_1$ and $C_{-1}$ cannot both be tuned to zero at the same time. Another way to say this is to note that in our choice of parameters, the resolution parameter $\alpha$ is determined in terms of $L$, with no dependence on $C_a$. It is not clear to us if this is a general feature of gauge transformations that allow resolution of the singular cosmology or a feature of the specific ansatzes that we chose.

In any event, the conclusion is that we have resolved the singularity in the quotient cosmology metric. We accomplished this by (effectively) doing a higher spin gauge transformation that preserved the holonomy and the flatness. %{\bf (Contradictory statements in Maloney's paper.)} 
Note however that the asymptotic geometry is no longer of the conventional asymptotically dS$_3$ form. A similar price had to be paid for resolving the black holes singularity in \cite{Maloney} where one ended up losing the conventional asymptotically AdS$_3$ form. This is to be expected: a spin-3 field is non-normalizable and corresponds to deforming the boundary theory. This is also reflected in the fact that the higher spin field components can vanish in the interior or reach the future or past boundary in our resolved solution. Similar features were again seen in \cite{Maloney}. To complete the presentation of our resolved version of the solution, we present these spin-3 fields in an appendix.

\section*{Acknowledgments}

CK thanks Justin David, Rajesh Gopakumar, Prem Kumar and Ashoke Sen for discussions, Ben Craps and Oleg Evnin for conversations on stringy big-bang singularity resolution back in 2008 in Cyprus, and the organizers and participants of the ICTS workshop at TIFR, Mumbai (Jan 7-9, 2013) for a stimulating discussion session on higher spin theories. This manuscript was %Alok Maharana, 
finalized while CK was visiting IISER, Mohali. He thanks Alok Maharana, Yashonidhi Pandey, Shyam Sundar and especially Kalpat Pattabhirama ``Patta" Yogendran for inspiring discussions (on math, physics, higher spins and other things) at Mohali and Shimla. The research of SR is supported by Department of Science and Technology (DST), Govt.$\hspace{0.05in}$of India research grant under scheme DSTO/1100 (ACAQFT). Last but most, we thank Avinash Raju and Somyadip Thakur for collaboration on a related paper \cite{HSCosmology}. 
\appendix
%dummy comment inserted by tex2lyx to ensure that this paragraph is not empty
%dummy comment inserted by tex2lyx to ensure that this paragraph is not empty

\section{$SL(2,\IC)$ gauge field in terms of (complex) Euler angles}

The $SL(2)$ generators we use are

\[
T_{0}=-i\left(\begin{array}{ccc}
1\\
 & 0\\
 &  & -1
\end{array}\right),\qquad T_{1}=\frac{i}{\sqrt{2}}\left(\begin{array}{ccc}
 & -1\\
1 & 0 & -1\\
 & 1
\end{array}\right),\qquad T_{2}=\frac{1}{\sqrt{2}}\left(\begin{array}{ccc}
 & 1\\
1 &  & 1\\
 & 1
\end{array}\right),
\]
 and the corresponding subgroups are, 
\begin{eqnarray*}
U_{0}=e^{\theta_{0}T_{0}} & = & I+\sin\theta_{0}\; T_{0}+2\sin^{2}\frac{\theta_{0}}{2}\; T_{0}^{2}\\
U_{1}=e^{\theta_{1}T_{1}} & = & I+\sinh\theta_{1}\: T_{1}+2\sinh^{2}\frac{\theta_{1}}{2}\: T_{1}^{2}\\
U_{2}=e^{\theta_{2}T_{2}} & = & I+\sinh\theta_{2}\: T_{2}+2\sinh^{2}\frac{\theta_{2}}{2}\; T_{2}^{2}.
\end{eqnarray*}
 For $U=U_{0}U_{2}U_{1}$, one has, 
\[
U^{-1}dU=\left(U_{2}U_{1}\right)^{-1}\left(U_{0}^{-1}dU_{0}\right)
\left(U_{2}U_{1}\right)+U_{1}^{-1}\left(U_{2}^{-1}dU_{2}\right)U_{1}+U_{1}^{-1}dU_{1}.
\]
%{\bf (Looks to me that the $U_1 U_2$ in the first term or should it be $U_2 U_1$?)}
 Using, 
\[
U_{1}^{-1}dU_{1}=d\theta_{1}T_{1},
\]
 
\[
U_{1}^{-1}\left(U_{2}^{-1}dU_{2}\right)U_{1}=\left(\cosh\theta_{1}T_{2}-\sinh\theta_{1}T_{0}\right)d\theta_{2}
\]
 and, 
\[
\left(U_{2}U_{1}\right)^{-1}\left(U_{0}^{-1}dU_{0}\right)\left(U_{2}U_{1}\right)=\left(\cosh\theta_{1}\cosh\theta_{2}T_{0}+\sinh\theta_{2}T_{1}-\sinh\theta_{1}\cosh\theta_{2}T_{2}\right)d\theta_{0}
\]
 we finally have the expression for a pure gauge field $A=U^{-1}dU$
\begin{eqnarray}
A^{0} & = & \cosh\theta_{1}\cosh\theta_{2}d\theta_{0}-\sinh\theta_{1}d\theta_{2}\nonumber \\
A^{1} & = & d\theta_{1}+\sinh\theta_{2}d\theta_{0}\nonumber \\
A^{2} & = & -\sinh\theta_{1}\cosh\theta_{2}d\theta_{0}+\cosh\theta_{1}d\theta_{2}\label{eq:gauge field components in terms of Euler angles}
\end{eqnarray}

\section{Wilson Loop in the Fefferman-Graham coordinates}

Proceeding identically as in the last section for $V=e^{\psi_{2}T_{2}}e^{\psi_{0}T_{0}}$,
we get, 
\[
V^{-1}dV=d\psi_{0}T_{0}-\sin\psi_{0}\: d\psi_{2}\: T_{1}+\cos\psi_{0}\: d\psi_{2}\: T_{2}.
\]
 with, 
\[
\psi_{0}=i\left(\tau-\frac{1}{2}\ln\frac{L}{2l}\right),\qquad\psi_{2}=i\sqrt{\frac{2L}{l}}\left(\phi+\frac{iz}{l}\right).
\]
 The corresponding Wilson loop in the $\phi$- direction, 
 %({\bf We surely mean $2 \pi$ in place of $2 \phi$ below?)}
\[
W=V^{-1}(z,\tau,0)V(z,\tau,2\pi)
\]
 is, 
\[
\left(\begin{array}{ccc}
\cosh^{2}\frac{\Delta\psi_{2}}{2} & \frac{e^{i\psi_{0}}\sinh\Delta\psi_{2}}{\sqrt{2}} & e^{2i\psi_{0}}\sinh^{2}\frac{\Delta\psi_{2}}{2}\\
\frac{e^{-i\psi_{0}}\sinh\Delta\psi_{2}}{\sqrt{2}} & \cosh\Delta\psi_{2} & \frac{e^{i\psi_{0}}\sinh\Delta\psi_{2}}{\sqrt{2}}\\
e^{-2i\psi_{0}}\sinh^{2}\frac{\Delta\psi_{2}}{2} & \frac{e^{-i\psi_{0}}\sinh\Delta\psi_{2}}{\sqrt{2}} & \cosh^{2}\frac{\Delta\psi_{2}}{2}
\end{array}\right)
\]
 where, 
\[
\Delta\psi_{2}=2\pi\, i\,\sqrt{\frac{2L}{l}}.
\]
 The Eigenvalues of this matrix are, 
\[
\left(1,e^{-\Delta\psi_{2}},e^{\Delta\psi_{2}}\right).
\]

\section{The SL(3) basis\label{sl3app}}

Consistent with \cite{Kraus1, Maloney}, we have employed the following
set of generators which we use to furnish a basis for $sl(3)$: %{\bf This is just $sl(3)$ not $sl(3,\IR)$, because what we are after is $SL(3,\IC)$}.
%\[ T_{0}=-i\left(\begin{array}{ccc} 1\\  & 0\\ &  & -1 \end{array}\right),\qquad T_{1}=\frac{i {\sqrt{2}}\left(\begin{array}{ccc} & -1\\ 1 & 0 & -1\\ & 1 \end{array}\right),\qquad T_{2}=\frac{1}{\sqrt{2}}\left(\begin{array}{ccc}  & 1\\ 1 &  & 1\\& 1\end{array}\right), \]
\[ L_{1}=\left(\begin{array}{ccc} 0 & 0 & 0\\  1 & 0& 0\\  0 &  1 & 0 \end{array}\right),\qquad L_{0}=\left(\begin{array}{ccc}  1& 0&0\\ 0 & 0 & 0\\  0& 0 &-1 \end{array}\right),\qquad L_{-1}=\left(\begin{array}{ccc} 0 & -2&0\\ 0 & 0 & -2\\ 0 & 0 &0 \end{array}\right), \]
\[
W_{1}=\left(\begin{array}{ccc}
0 & 0 & 0\\
1 & 0 & 0\\
0 & -1 & 0
\end{array}\right),\qquad W_{0}=\left(\begin{array}{ccc}
2/3 & 0 & 0\\
0 & -4/3 & 0\\
0 & 0 & 2/3
\end{array}\right),\qquad W_{-1}=\left(\begin{array}{ccc}
0 & -2 & 0\\
0 & 0 & 2\\
0 & 0 & 0
\end{array}\right)
\]
 
\[
W_{2}=\left(\begin{array}{ccc}
0 & 0 & 0\\
0 & 0 & 0\\
2 & 0 & 0
\end{array}\right),\qquad W_{-2}=\left(\begin{array}{ccc}
0 & 0 & 8\\
0 & 0 & 0\\
0 & 0 & 0
\end{array}\right)
\]
Note that we have set $\sigma=-1$ in \cite{Kraus1}.
The algebra they satisfy is
\bea
&&[L_m,L_n] = (m-n) L_{m+n}\\
&&[L_m,W_p] = (2m-p) W_{m+p} \\
&&[W_p,W_q] = \frac{\sigma}{3}(p-q) (2p^2 + 2q^2 -pq-8) L_{p+q}. \label{lastww}
\eea
Using these we define the $T_a$ as in (\ref{TintermsofL}) and the $SL(3)$ algebra takes the following form:
\[
\left[T_{a},T_{b}\right]=\epsilon_{ab}\,^{c}T_{c},
\]
\[
\left[T_{0},W_{a}\right]=i\, a\, W_{a},
\]
and the algebra between the $W$'s is of course the same as (\ref{lastww}).

%The corresponding group elements,  \[e^{\phi_{1}W_{1}}=I+\phi_{1}W_{1}-\frac{1}{2}\frac{\phi_{1}^{2}}{2!}W_{2},\qquad e^{\phi_{-1}W_{-1}}=I+\phi_{-1}W_{-1}-\frac{1}{2}\frac{\phi_{-1}^{2}}{2!}W_{-2}, \] \[ e^{\phi_{0}W_{0}}=\left(\begin{array}{ccc} e^{2\phi_{0}/3}\\ & e^{-4\phi_{0}/3}\\ &  & e^{2\phi_{0}/3} \end{array}\right),\] \[ e^{\phi_{2}W_{2}}=I+\phi_{2}W_{2},\qquad e^{\phi_{-2}W_{-2}}=I+\phi_{-2}W_{-2}. \] and

\section{Flatness of the Primitive $SL(3,\IC)$ connection %$A'$
\label{sub:Flatness condition}}

%{\bf I changed $SL(3,\IR)$ to $SL(3,\IC)$ in the title of this subsection.)}

In this appendix we will show that the primitive connection $a$ of the type (\ref{firsta}) or (\ref{seconda}) that we use to construct our solutions are flat.
%in last Sec. \ref{sec:General Case}. 
By directly plugging in the matrix-valued one-form
$a=\left(a_{\mu}^{i}L_{a}+a_{\mu}^{m}W_{m}\right)dx^{\mu}$, the field strength 
\bea
F_{\mu\nu}=\partial_{\mu}a_{\nu}-\partial_{\nu}a_{\mu}+\left[a_{\mu},a_{\nu}\right]
\eea
can be explicitly computed
%\[
%\left(dA\right)_{\mu\nu}=\partial_{\mu}A_{\nu}-\partial_{\nu}A_{\mu}=
%\left(\partial_{\mu}A_{\nu}^{i}-\partial_{\nu}A_{\mu}^{i}\right)L_{i}+\left(\partial_{\mu}A_{\nu}^{m}-\partial_{\nu}A_{\mu}^{m}\right)W_{m}
%\]
using the $SL(3)$ algebra relations in \cite{Maloney} (or our Appendix \ref{sl3app}). 
%\begin{align*}
%\left(A\wedge A\right)_{\mu\nu} & %=\left[A_{\mu},A_{\nu}\right]=A_{\mu}^{i}A_{\nu}^{j}(i-j)L_{i+j}+\left(A_{\mu}^{i}A_{\nu}^{m}-A_{\mu}^{m}A_{\nu}^{i}\right)(2i-m)W_{i+m}\\
% & \qquad\qquad\qquad\qquad\qquad\qquad\qquad\qquad-\frac{1}{3}A_{\mu}^{m}A_{\nu}^{n}(m-n)\left(2m^{2}+2n^{2}-mn-8\right)L_{m+n}
%\end{align*}
%The equation of motion is,
%\[
%dA+A\wedge A=0
%\]
The $w\bar{w}$ and the $\rho\bar{w}$ components of the equation
of motion are manifestly satisfied (due to the fact that every component
is at best a function of $\rho$ and in our gauge $a_{\bar{w}},a_{\rho}=0$).
The $\rho w$ component equation is,
\[
\partial_{\rho}a_{w}=0
\]
so the components are forced to be $\rho$-independent. %\textbf{
\[
\implies C_{a}=\mbox{constant}
\]
%}
So this is the condition for a connection like $a$ to
be flat. Then it follows, that $A$=$b^{-1}a b+b^{-1}db$, 
being a gauge transform of $a$ will also flat. A similar statement also holds for $A'$. 

Even though we don't look at more general gauge fields, it is worthwhile mentioning that $C_a$ can be arbitrary functions of $w$, and the connection will still be flat. 

\section{The Spin-3 Field on the Resolved Geometry}

For the connection in Eq.(\ref{eq: regular SL(3)}) obtained by performing
a spin-3 transformation on a singular pure $SL(2,\mathbb{C})$ geometry,
one turns on the spin-3 field (following conventions of \cite{Campo}),
\[
\phi_{\mu\nu\rho}=\frac{1}{3!}\mbox{Tr}\left(e'_{(\mu}e'_{\nu}e'_{\rho)}\right)
\]
where
\begin{align*}
e' & =e+\frac{Xdw-\bar{X}d\bar{w}}{2i/l}\\
 & =e^{a}T_{a}-\frac{il}{2}\sum_{a=-2}^{2}e^{a\tau}C_{a}W_{a}dw+\frac{il}{2}\sum_{a=-2}^{2}e^{a\tau}\bar{C}_{a}W_{a}d\bar{w.}
\end{align*}
Explicitly
\[
e'_{w}=-\frac{i\, l}{2}\left(e^{\tau}-\frac{L}{2l}e^{-\tau}\right)\, T_{1}+\frac{l}{2}\left(e^{\tau}+\frac{L}{2l}e^{-\tau}\right)\, T_{2}-\frac{il}{2}\sum_{a=-2}^{2}e^{a\tau}\, C_{a}\: W_{a}
\]
\begin{align*}
e'_{\bar{w}} & =\frac{i\, l}{2}\left(e^{\tau}-\frac{\bar{L}}{2l}e^{-\tau}\right)\, T_{1}+\frac{l}{2}\left(e^{\tau}+\frac{\bar{L}}{2l}e^{-\tau}\right)\, T_{2}+\frac{il}{2}\sum_{a=-2}^{2}e^{a\tau}\:\bar{C}_{a}\: W_{a}\\
e'_{\tau} & =e_{\tau}=l\, T_{0},
\end{align*}
which then determine the components of the spin-3 field. For the case where we have chosen the parameters $C_a$ as in (\ref{eq:Regular metric}) the result is
\[
\phi_{\tau\tau\tau}=\phi_{www}=\phi_{\bar{w}\bar{w}\bar{w}}=0,
\]
\[
\phi_{\tau\tau w}=\phi_{\tau\tau\bar{w}}=-\frac{\sqrt{2}l^{3}}{6}\sqrt{\frac{L}{l}},
\]
\[
\phi_{ww\tau}=\frac{l^{2}(4C_{-1}l-C_{1}L)}{12\sqrt{2}},\phi_{ww\bar{w}}=-\frac{\sqrt{Ll}\left(4e^{2\tau}l^{2}+20Ll+
e^{-2\tau}L^{2}\right)}{24\sqrt{2}}
\]
\[
\phi_{\bar{w}\bar{w}\tau}=\frac{l^{2}\left(e^{2\tau}C_{1}l-e^{-2\tau}C_{-1}L\right)}{12\sqrt{2}},\phi_{\bar{w}\bar{w}w}=-\frac{\sqrt{Ll}\left(4e^{2\tau}l^{2}
+20Ll+e^{-2\tau}L^{2}\right)}{24\sqrt{2}}
\]
\[
\phi_{\tau w\bar{w}}=\frac{l^{2}\left(C_{1}+2e^{-2\tau}C_{-1}\right)
\left(L-2e^{2\tau}l\right)}{24\sqrt{2}}.
\]
Here $C_{1}$ and $C_{-1}$ could be any pair of real numbers satisfying,
\[
C_{1}C_{-1}=-\frac{3L}{2l}.
\]

An observation worthy of remark here is that some components of the spin-3 fields vanish and some reach all the way to the (future or past) boundary. This means that the higher spin gauge transformations that we used have come at a price: a similar phenomenon was observed in the context of higher spin black holes in AdS$_3$. %singularities in the spin-3 field while removing the singularities in the metric.

%\appendix
%dummy comment inserted by tex2lyx to ensure that this paragraph is not empty
%dummy comment inserted by tex2lyx to ensure that this paragraph is not empty
% \bibliographystyle{brownphys}
%\bibliography{Kerr-dS3}
%%%%%%%%%%%%%%%%%%%%%%%%%%%%%%%%%%%%%%%%%%%%%%%%%%%%%%%%%%%%%%%%%%%%%%%%%%%%%%%%%%%%%%%%%%%%%%%

%%%%%%%%%%%%%%%%%%%%%%%%%%%%%%%%%%%%%%%%%%%%%%%%%%%%%%%%%%%%%%%%%%%%%%%%%%%%%%%%
%%%%%%%%%%%%%%%%%%%%%%%%%%%%%%%%%%%%%%%%%%%%%%%%%%%%%%%%%%%%%%%%%%%%%%%%%%%%%%%%

%%%%%%%%%%%%%%%%%%%%%%%%%%%%%%%%%%%%%%%%%%%%%%%%%%%%%%%%%%%%%%%%%%%%%%%%%%%%%%%%%%%%%%%%%%%%%%%%%%%%%%

% ==========================================================================
%
%%%%%%%%%%%%%%%%%%%%%%%%%%%%%%%%%%%%%%%%%%%%%%%%%%%%%%%%%%%%%%%%%%%%%%%%%%%%
%                      REFERENCES                            %
%%%%%%%%%%%%%%%%%%%%%%%%%%%%%%%%%%%%%%%%%%%%%%%%%%%%%%%%%%%%%%%%%%%%%%%%%%%%
%\newpage
%\bibliography{metasusy}

\end{document}